\documentclass[conference]{IEEEtran}
\usepackage{cite}
\usepackage{adjustbox}
\usepackage{graphicx,lineno}
\usepackage{algorithm}
\usepackage{algorithmic}
\usepackage{multirow}
\usepackage[tight,footnotesize]{subfigure}
\usepackage{epsfig}
\usepackage{amssymb}
\usepackage{footmisc}
\usepackage{footnpag}
\usepackage{amsmath}
\usepackage{paralist}
\usepackage{color}
\usepackage{colortbl}
\usepackage{perpage}
\usepackage{fixltx2e}
\usepackage{verbatim}
\usepackage{comment} 
\usepackage{eqparbox}
\usepackage{array}
\usepackage{enumitem}
\usepackage{balance}

\usepackage{tikz}

\newlength\savedwidth
\newcommand\whline{\noalign{\global\savedwidth\arrayrulewidth
                            \global\arrayrulewidth 1.5pt}%
           \hline
           \noalign{\global\arrayrulewidth\savedwidth}}

\newcommand*\circled[1]{\tikz[baseline=(char.base)]{
            \node[shape=circle,draw,inner sep=0.8pt] (char) {#1};}}

\usepackage{color}

\newcommand{\Comment}[2]{{\color{red} Comment by {#1:\ }{#2}}}
\newcommand{\noComment}[2]{{}}
\newcommand{\pscale}{\alpha}
\newcommand{\pbias}{\beta}
\newcommand{\scale}[1]{\pscale_{{#1}}}
\newcommand{\bias}[1]{\pbias_{{#1}}}
\newcommand{\indset}{{\cal I}}

\begin{document}


\title{Efficient Soft-Error Detection for Low-precision Deep Learning Recommendation Models}

\author{\IEEEauthorblockN{Sihuan Li,\IEEEauthorrefmark{1}
Jianyu Huang,\IEEEauthorrefmark{1}
Ping Tak Peter Tang,\IEEEauthorrefmark{1} Daya Khudia,\IEEEauthorrefmark{1} Jongsoo Park,\IEEEauthorrefmark{1} \\Harish Dattatraya Dixit,\IEEEauthorrefmark{1} and
Zizhong Chen\IEEEauthorrefmark{2}}
\IEEEauthorblockA{\IEEEauthorrefmark{1}
Facebook, Inc.
}
\IEEEauthorblockA{\IEEEauthorrefmark{2}
University of California, Riverside, CA, USA
}

\{sihuan, jianyuhuang, ptpt, dskhudia, jongsoo, hdd\}@fb.com, chen@cs.ucr.edu
}

\pagestyle{plain}


\maketitle
\begin{abstract}
Soft error, namely silent corruption of signal or datum in a computer system, cannot be caverlierly ignored as compute and communication density grow exponentially. Soft error detection has been studied in the context of enterprise computing,  high-performance computing and more recently in convolutional neural networks related to autonomous driving. 

Deep learning recommendation systems (DLRMs) have by now become ubiquitous and serve billions of users per day. Nevertheless, DLRM-specific soft error detection methods are hitherto missing. To fill the gap, this paper presents the first set of soft-error detection methods for low-precision quantized-arithmetic operators in DLRM including general matrix multiplication (GEMM) and EmbeddingBag. A practical method must detect error and do so with low overhead lest reduced inference speed degrades user experience. Exploiting the characteristics of both quantized arithmetic and the operators, we achieved more than 95\% detection accuracy for GEMM with an overhead below 20\%. For EmbeddingBag, we achieved 99\% effectiveness in significant-bit-flips detection with less than 10\% of false positives, while keeping overhead below 26\%.

\end{abstract}

\section{Introduction}
\label{sec: intro}

Hardware faults can be separated into two categories: fail-stop and fail-continue.
Fail-stop faults crash the executing process, thus detectable by the operating system, and can be handled by well-studied checkpoint-and-restart techniques~\cite{checkpoint-di}. In contrast, fail-continue faults silently corrupt the results of an execution process without interrupting it. The induced errors are usually called soft errors~\cite{soft-errors} or silent data corruptions and are the focus of this paper.

Soft error is much more prevalent than one may realize: even experienced practitioners grossly underestimate their frequency of occurrences~\cite{geist2016kill}. 
The supercomputer Jaguar, for example, suffers a double-bit memory error once every 24 hours \cite{geist2016scmonster}. {For another example, a recent research study taking more than 18 months \cite{dixit2021silent} has confirmed the large-scale infrastructure at Facebook is experiencing silent data corruptions due to device characteristics inside hundreds of Central Processing Units (CPUs).} The situation is only worsening: not only cosmic radiation triggers soft error but simple down-to-earth factors such as temperature and power consumption can also be the culprit~\cite{nie2017characterizing}.
Further exacerbating the situation is the rapid emergence of deep-learning ASIC accelerators which are prone to have higher error rates than general-purpose computing hardware do \cite{zhang2018thundervolt}.

Enterprise computing is the first to employ soft error detection, followed by the HPC community~\cite{wu2017silent,chen2018fault,liang2017correcting,chen2013online,tao2016new}, and most recently, error detection methods are explored for convolutional neural networks (CNNs) deployed in autonomous driving~\cite{abftcnngpu2020making}. While deep learning recommendation systems (DLRMs) may not be critical to personal safety, their computational integrity is crucial to maintain good experience of billions of users per day. A deployable soft-error detection method for DLRMs must only incur low performance overhead lest the goal of maintaining user experience be self defeated, making algorithmic based fault tolerant method (ABFT)~\cite{huang1984algorithm} the prime candidate. However, to the best of our knowledge, there is no previous ABFT work targeting DLRMs which typically compute in low-precision quantized arithmetic (details in~\ref{sec: quantized arith}).

The two workhorse operators of DLRM are general matrix-matrix multiplication (GEMM) and EmbeddingBag (EB) which together account for over 70\% of a DLRM's compute latency. Although ABFT for GEMM has been well studied in the literature, their straightforward adaption to DRLM results in high overhead due to DLRM's peculiar matrix sizes and shapes and its use of quantized arithmetic. In addition, EB is an operator not present in HPC or even in convolutional neural network. This paper considers error detection by ABFT for these two operators.
We do not focus on error resilience as that is relatively simple for recommendation systems: once an error is detected a recommendation score can be recomputed easily assuming error striking twice is very rare.

The paper makes the following contributions on efficient soft-error detection for the key building blocks of DLRM in the quantized arithmetic domain:
\begin{itemize}
    \item We propose the first ABFT implementation for quantized GEMM. By carefully customizing ABFT for GEMM, we optimize the performance and analyze its error detection ability;
    \item We propose the first ABFT implementation for EB, which is especially important for recommendation models.
\end{itemize}

In the following, we first briefly review related works in Section~\ref{sec: related work}, followed by Section~\ref{sec:background} which explains the low-precision arithmetic used in many industrial machine learning models including DLRMs, and its two main operators that we focus on. Sections~\ref{sec: abft for gemm} and \ref{sec: abft for eb} present our two ABFT algorithms and implementation considerations. Section~\ref{sec:evaluation} presents our experiments to support our statements on the proposed algorithms' performance and efficacy. Section~\ref{sec:conclusion} makes some concluding remarks.


\section{Related Work} \label{sec: related work}
Soft error resiliency in deep learning models have been attracting more and more attention in recent years. Redundancy based protections are the most general and reliable solutions, where redundancy can be done at the hardware or software level. Hardware level redundancy is usually used in safety-critical task such as self-driving~\cite{tesla2020compute}. Software level redundancy can be done in the same hardware but with duplicated or tripled program or instruction executions \cite{reis2005swift}. Error detection by redundancy incurs at least 100\% in overhead. 

ABFT is a low-overhead error detection method. Though less general than using redundancy, it is shown to be effective on convolutional neural networks (CNN)~\cite{abftgemmforconv2018analyzing, abftcnngpu2020making,zhao2020algorithm}. These ABFT works either target convolution specifically or rely on extra-precision intermediate computation. While one can adapt to some extent these work to GEMM, we aim to eliminate the use of extra precision to further reduce overhead and devise ABFT for GEMM that is DLRM-specific. Furthermore, the EB operator is peculiar to DLRMs and hitherto unexplored. 

\section{Arithmetic and Operators}
\label{sec:background}
\subsection{Quantized Arithmetic}
\label{sec: quantized arith}
Deep learning intrinsically relies on computing with real numbers. If the representation and computation of these real numbers can use, say, 8-bit integers instead of 32-bit floating-point numbers, significant memory saving and performance boost can be obtained. Arithmetic in integer to approximate floating-point computation is commonly called quantized arithmetic~\cite{jacob2018quantization}. One first transforms linearly an interval $[x_{\rm min},x_{\rm max}]$ of interest to the domain of the integer arithmetic in question. For example, $[0,255]$ for 8-bit unsigned integer: Determine floating-point numbers $\pscale$, $\pbias$ so that $(x - \pbias)/\pscale \in [0,255]$ for all $x\in [x_{\rm min},x_{\rm max}]$. The resulting value is then rounded to an integer, yielding $x_I$, hence $x \approx \pscale x_I + \pbias$. In quantized arithmetic, instead of multiplying two floating-point matrices $A\times B$ of dimension $m\times k$ and $k\times n$, the matrices are represented by $(A_I,\scale{A},\bias{A})$ and $(B_I,\scale{B},\bias{B})$ and the corresponding matrix product is realized as integer matrix product:
\begin{align}\label{eqn: quantized matrix product}
AB &\approx (\scale{A}A_I + \bias{A}\vec{e}_m\vec{e}_k^T) (\scale{B}B_I + \bias{B}\vec{e}_k\vec{e}_n^T) \nonumber \\
&= \scale{A}\scale{B}A_I B_I + \nonumber \\
&\phantom{=}\scale{A}\bias{B}(A_I\vec{e}_k)\vec{e}_n^T +
\scale{B}\bias{A}\vec{e}_m(\vec{e}_k^T B_I) + k\bias{A}\bias{B}\vec{e}_m\vec{e}_n^T
\end{align}
where $\vec{e}_\ell$ is the dimension-$\ell$ vector of all ones. Note all the terms following $A_I B_I$ are rank-1 matrices. 

\subsection{Low-precision GEMM in DLRM}
\begin{figure}[h]
    \centering
    \includegraphics[width=\columnwidth]{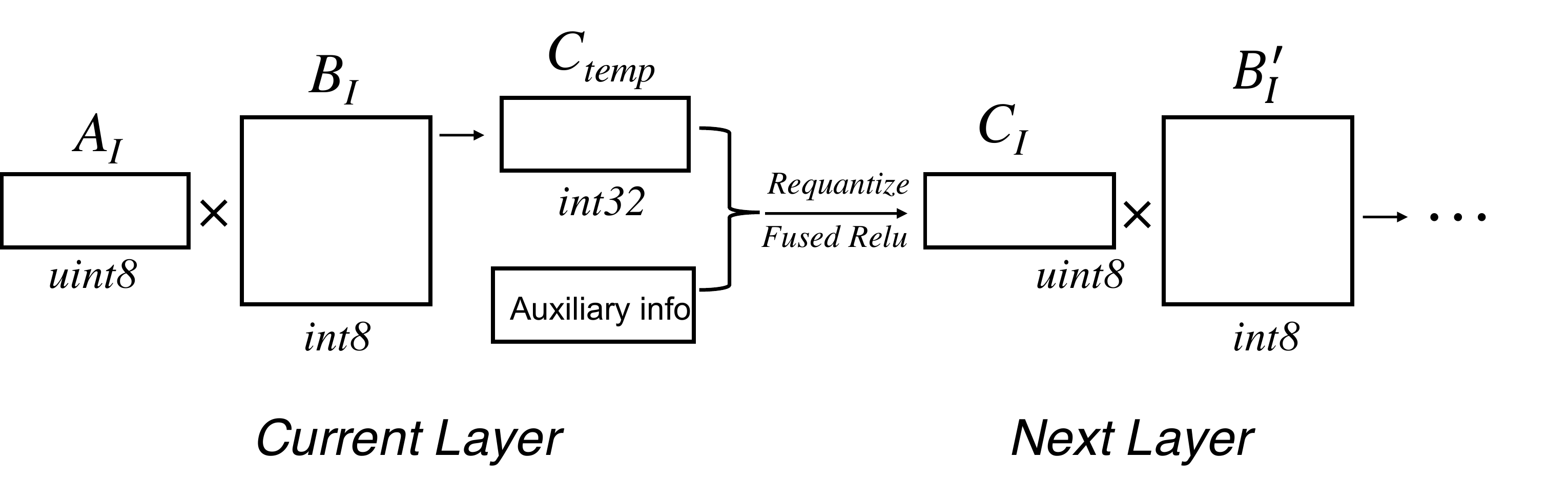}
    \caption{Low-precision GEMM in deep learning context}
    \label{fig: low precision gemm}
\end{figure}

Industrial implementations of DLRMs exploiting quantized arithmetic typically use specialized high-performance libraries such as FBGEMM~\cite{khudia2021fbgemm}. As shown in Equation~\ref{eqn: quantized matrix product}, the dominant operation is the integer matrix product $C_{\rm temp} = A_I B_I$, consisting of $2mnk$ operations. This $C_{\rm temp}$, a 32-bit integer matrix, together with the other rank-1 matrices and miscellaneous scale factors are than  combined in a requantization process producing the $C \approx A B$ where $C$ is represented by the tuple $(C_I, \scale{C}, \bias{C})$. We show the workflow in Figure~\ref{fig: low precision gemm}. In the rest of the paper, when we refer to the matrices using $A$, $B$, $C$, they are corresponded to integer matrices $A_I$, $B_I$, $C_{\rm temp}$ for notation simplicity.

\subsection{EmbeddingBag and its low-precision variant}
\label{sec: eb background}
\begin{figure}[h]
    \centering
    \includegraphics[width=0.7\columnwidth]{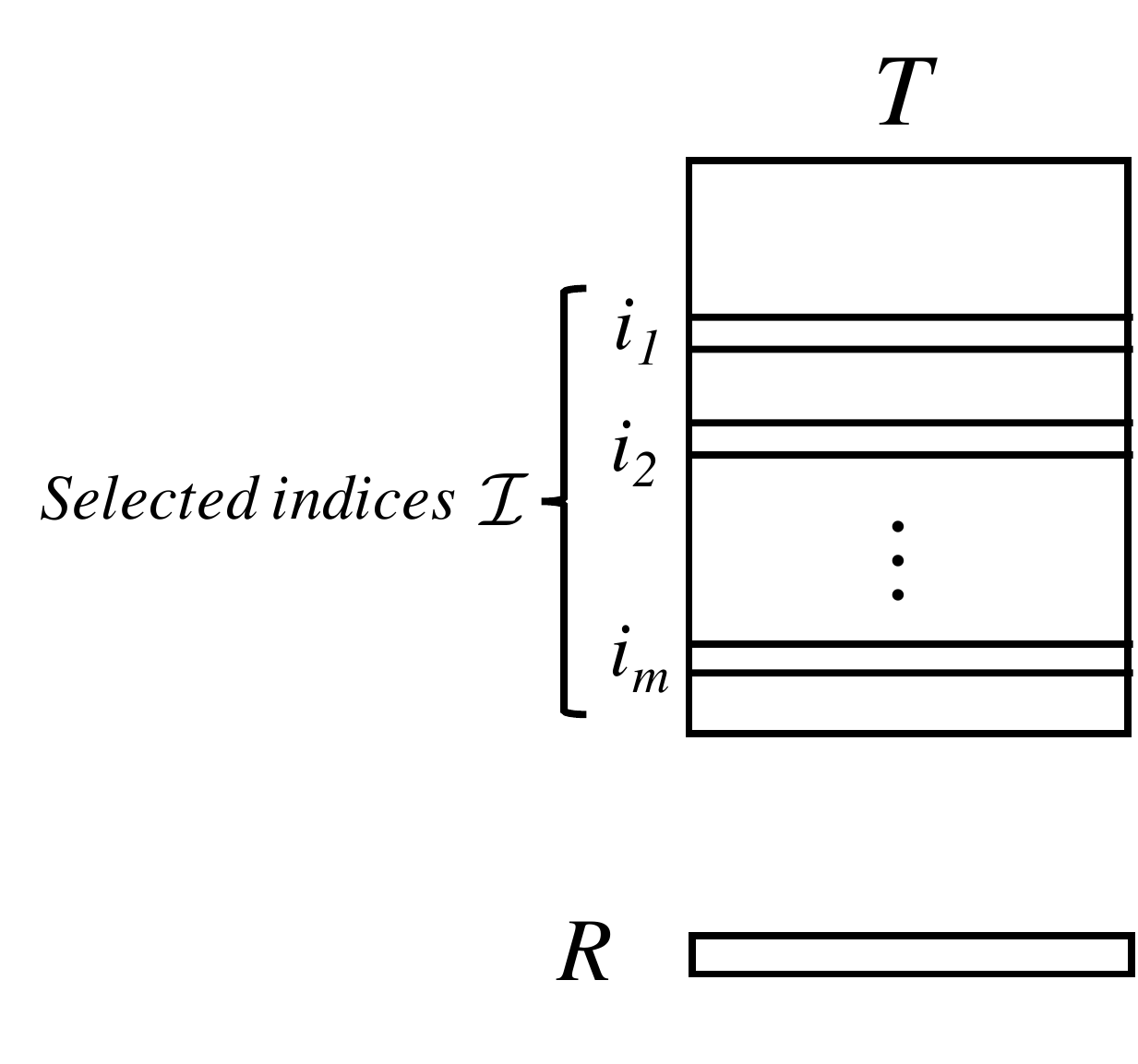}
    \caption{Illustration of one embedding lookup in the EmbeddingBag operator with batch size of one}
    \label{fig: eb operation}
\end{figure}

Embedding is a technique that maps discrete categorical data into a $d$-dimensional Euclidean spaces of real numbers. It is widely used in many recommendation systems \cite{zhao2019aibox, zhao2020distributed, xie2020kraken}. An embedding table contains a number of $d$-length row vectors each corresponding to a categorical data and algebraic operations corresponds to combination of these categories. EmbeddingBag (EB)\footnote{https://pytorch.org/docs/stable/generated/torch.nn.EmbeddingBag.html} is one of the most frequently called operators in these embedding based recommendation systems. An EB with batch size of one simply picks out the set of rows given by an index set $\indset$ from an embedding table and sum them up, illustrated in Figure~\ref{fig: eb operation}. It is also called one embedding lookup. Mathematically, given $\indset$ and an embedding table $T$, EB returns $\vec{R} = \sum_{i\in\indset} \vec{eb}_i$, where $\vec{eb}_i$ is the $i$-th row of the embedding table $T$. Note that for notational convenience we use $\vec{r}$ here to denote a \emph{row} vector instead of the usual convention of a column vector.

Industrial scale DLRMs often have many embedding tables totaling hundred billions of parameters. Hence instead of using floating-point to represent these real numbers, quantized arithmetic is often used to reduce the DLRMs' memory footprints~\cite{guan2019four-bit}. Specifically, each $d$-length embedding row at index $i$ is represented by $d$-length vector of short (8 bits for example) integers $\vec{eb}_i$ and one pair of floating-point quantization parameter $\scale{i}, \bias{i}$. The corresponding EB operator must then compute $\vec{R} = \sum_{i\in\indset}\scale{i}\vec{eb}_i + \bias{i}\vec{e}_d$ where $\vec{e}_d$ is a $d$-length row vector of all ones.

As EB with batch size of one returns one row vector, EB with batch size $n$ returns $n$ vectors, each corresponding to sums of relevantly selected embedding rows from a particular embedding table:
\begin{align}
    R &= \begin{bmatrix}
           \vec{R}_1 \\
           \vec{R}_2 \\
           \vdots \\
           \vec{R}_n
         \end{bmatrix}
  \end{align}

\section{Optimized ABFT for GEMM in DLRMs}
\label{sec: abft for gemm}

The bulk of the computation in a quantized matrix product (Equation~\ref{eqn: quantized matrix product}) is the usual matrix product of two integer matrices of the form
$C = A B$ (dropping the subscripts of $I$) where $A$ and $B$ are matrices of dimensions $m$-by-$k$ and $k$-by-$n$, respectively. Our aim is to detect soft error that happen during this computation after both $A$ and $B$ have been loaded into memory.

\begin{figure}[h]
    \centering
    \includegraphics[width=\columnwidth]{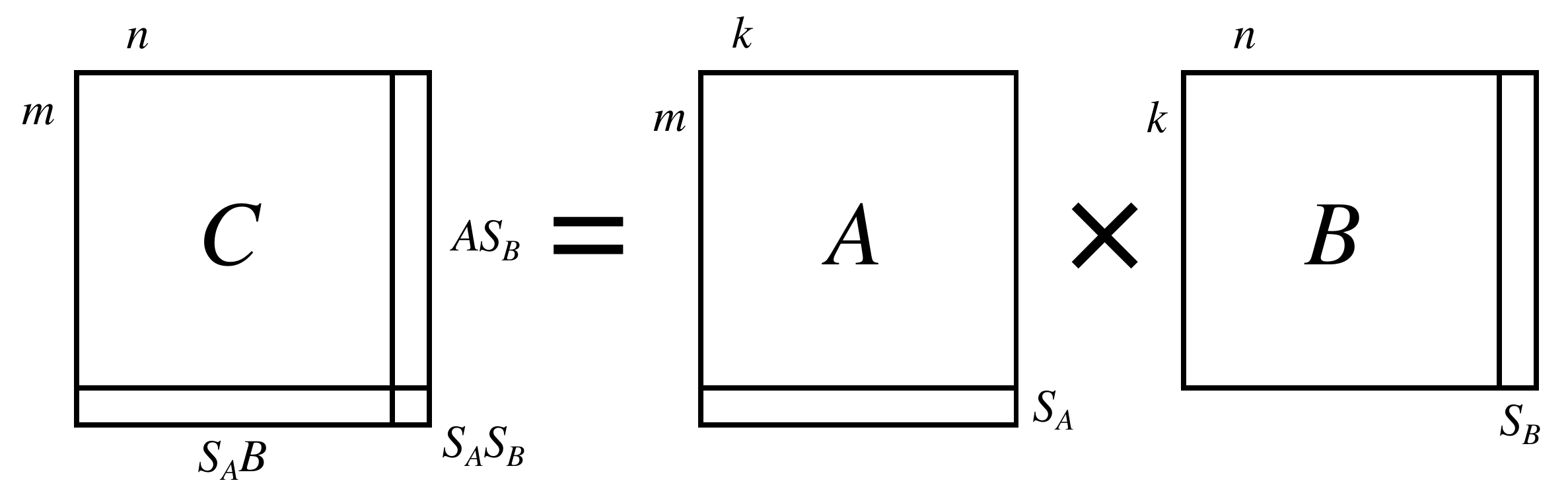}
    \caption{Illustration of ABFT for GEMM}
    \label{fig: abftgemm}
\end{figure}

We start with this common ABFT method: One encodes $A$ into an augmented matrix $A'$ by appending a row vector $S_A$ where $S_A[j] = \sum_{i=0}^{m-1} A[i][j]$. Similarly, $B$ is encoded into an augmented matrix $B'$ with an extra column vector $S_B$ where $S_B[i] = \sum_{j=0}^{n-1} B[i][j]$. Figure~\ref{fig: abftgemm} illustrates the augmented matrices $A'$ and $B'$ and their product $C'$. 
Mathematically, the upper-left $m$-by-$n$ block of $C'$ is $C = AB$, first $n$ columns of $C'[m,:]$ is
$S_A\,B$, first $m$ rows of $C'[:,n]$ is $S_A B$, and $C'[m,n] = S_A S_B$.
Simple algebraic derivations show that a correctly computed $C'$ satisfies the relationships
\begin{subequations}
\begin{equation} \label{eqt: column_chk}
    \forall {\rm column}\ j \in [0, n-1], C'[m][j] = \sum_{i=0}^{m-1} C'[i][j]
\end{equation}
\begin{equation} \label{eqt: row_chk}
    \forall {\rm row}\ i \in [0, m-1], C'[i][n] = \sum_{j=0}^{n-1} C'[i][j]
\end{equation}
\label{eqt: abft check}
\end{subequations}

Equality checks of these equations on the computed $C'$ form the basis of ABFT: If equality fails to hold at exactly one row $i$ for Equation~\ref{eqt: row_chk} together with exactly one column $j$ for Equation~\ref{eqt: column_chk}, then the value at the computed $C'[i][j]$ is faulty. Furthermore, a corrupted $C'[i][j]$ -- revealed as a single violation at row $i$ and at column $j$ -- can be corrected using the equation 
$${\rm correct}\ C'[i][j] = C'[m][j] - \sum_{p \neq i}C'[p][j],
$$
or
$$
{\rm correct}\ C'[i][j] = C'[i][n] - \sum_{p \neq j} C'[i][p].
$$
Straightforward as this common ABFT method for GEMM is, adopting them with low enough overhead that does not impede DLRM user experience requires a number of techniques that we now discuss.


\subsection{Performance optimizations}
\subsubsection{Encoding only matrix B}
Existing work of ABFT for GEMM considers soft error detection and single error correction. 
We stated previously (Section~\ref{sec: intro}) that
we aim solely at error detection. Thus, we just need to encode A or B, but not both. 
The question is which matrix to encode. To better understand this question, we first derive the theoretical error detection overheads with encoding A or encoding B. Remember that the error detection includes basically 3 stages: encode matrix A (or B); do GEMM with encoded A (or B); verify result matrix by checking Equation~\ref{eqt: abft check}. The overheads are:

\begin{align*}
\text{overhead if encoding $A$: }\frac{mk + 2nk + mn}{2mnk} &= \frac{1}{2n} + \frac{1}{m} + \frac{1}{2k} \\
\text{overhead if encoding $B$: }\frac{kn + 2mk + mn}{2mnk} &= \frac{1}{2m} + \frac{1}{n} + \frac{1}{2k} \\
\end{align*}

We follow the convention in PyTorch where $A$ corresponds to activations and $B$ the weight parameters of the neural network. Common in DLRMs, $m$ is relatively much smaller than $n$ or $k$. According to the theoretical overhead equation, encoding matrix $B$ will have smaller overheads than encoding $A$.

Another fact also makes encoding $B$ preferable in the aspects of both performance and memory error detection ability: $B$, being the trained weight matrix, stays still in the memory for a much longer time. From the perspective of performance overheads, the fact implies we can encode matrix $B$ once for multiple GEMM operations thus amortizes the encoding overheads. From the perspective of memory error detection ability, the fact implies matrix $B$ have much higher chances to experience memory errors than matrix $A$. (Recall that encoding matrix $A$ will not detect memory errors in $B$ while encoding matrix $B$ will do. In order to cover the errors in $B$, we choose to encode B.)
In conclusion, we encode $B$ instead of $A$ so as to minimize ABFT performance overheads while maximizing detection ability.

\subsubsection{Keeping encoded column in low precision}

The encoded row sum vector for matrix $B$ seems to require 32-bit integer as value container to ensure correctness. This implies a high overhead because ABFT work has to be done in 32-bit integers while the original GEMM work is done in 8-bit integers. Computation with 32-bit integers can be 2 to 4 times slower than that with 8-bit. To reduce the overheads, we use modulo operations to map the 32-bit row sums into 8-bit. The Equations~\ref{eqt: abft check} are proved to still hold under the same modulus \cite{huang1984algorithm}. Using modulo operations in the ABFT context is not novel. But we exploit them for better performance rather than to bypass limitation of computer word length~\cite{huang1984algorithm}.

\subsubsection{Keeping BLAS level-3 updating}
A straightforward implementation of ABFT for GEMM (encoding only $B$) will be: \circled{1} calculate row sums of $B$ and store the result ($S_B$) in a separate vector; \circled{2} compute $C = A*B$; \circled{3} compute $A*S_B$; \circled{4} check if row sums of $C$ equal $A*S_B$. This implementation does not need to modify the normal data structure of $B$ to accommodate an extra column, but results in high performance overhead. This is because Step \circled{3} is a matrix-vector product, a BLAS (Basic Linear Algebra Subprograms) level 2 operation. An alternative implementation that relies BLAS level 3 operations can be: \circled{1} allocate new memory for encoded matrix $B'$ and new memory for $C'$; \circled{2} do GEMM between $A$ and $B'$ and store result in $C'$; \circled{3} check Equation~\ref{eqt: abft check}; \circled{4} copy former $m$ rows and $n$ columns of $C'$ back into $C$. The drawback of this implementation is its high memory overhead.

We found a way to implement ABFT for low-precision GEMMs in BLAS level 3 operations and with small memory overhead. This is possible because of two facts: 1. matrix $B$ is packed into blocks before being sent to the efficient GEMM kernel; 2. the $C$ matrix in 32-bit integers are intermediate result (as shown in Figure~\ref{fig: low precision gemm} by $C_{\rm temp}$). The first fact means that we can pack the original $B$ and the separate vector storing row sums together into blocks so that the blocks look like they are from encoded $B'$ in contiguous memory space. The second fact means that we can directly allocate one more column for the intermediate result matrix than before. Notice we are not increasing the number of columns of 8-bit result matrix. We just need to modify the requantization procedure to let it exclude the last column of the intermediate 32-bit matrix.

\subsection{ABFT detection before requantization}
One may ask if we can delay the checksum equality check from examining $C_{\rm temp}$ (Equation~\ref{eqt: abft check}) to examining $C_I$ so as to detect silent errors in requantization process. Unfortunately, the answer is no. The main reason is that requantization is not a linear operation, i.e., $Q(a) + Q(b) \neq Q(a+b)$ generally where $Q$ is the requantization operator. Thus, our linear encoding scheme cannot make equality hold in $C_{I}$. Lack of error detection for requantization process is not serious considering that this process is less error prone as it only takes around 2\% of execution time for larger matrices and around 5\% for smaller matrices.


\subsection{{Modulus selection and detection ability analysis}}
\label{sec: detection ablt analysis}

As we use modulo to keep the encoded column in low precision to reduce ABFT overhead, the downside is the weakened error detection ability. In this section, we want to discuss how to choose the modulus wisely so that the detection ability degradation is minimized. We assume elements in 8-bit unsigned integer matrix A and 8-bit signed integer matrix B are both random numbers in the uniform distribution independently. We also assume the there are no errors for the encoded column considering its much smaller memory usage and operations numbers compared to the original computation. First, let us look at the situations when the modulus, $mod$, will fail to detect errors. For some rows in the result matrix $C$, we denote its row sum (excluding the encoded checksum column) by $rsum$ without any soft error. If soft errors happen and corrupt that row, we denote its row sum by $rsum'$. Then there is a fact that when the absolute value of difference between $rsum'$ and $rsum$ is divisible by $mod$, the soft error will not be detected. Also, that is only the case when the errors are not detected. More formally, soft errors corrupting that row are not detected if and only if $|rsum' - rsum| \% mod = 0$.

We consider two fault models. The first commonly used fault model is the random single-bit flip model which means a random bit of the data in the memory or register flips from 0 to 1 or 1 to 0. The intuition to this model is that $|rsum' - rsum|$ will be powers of 2. That is, any odd modulus can detect all errors in this model. The second model is random data fluctuation which means the correct value of the data is changed to some arbitrary value representable in its data type. 
For example, a 32-bit signed integer data can be changed to any value in the range of $[-2^{31}, 2^{31}-1]$ with equal likelihood. The intuition to this model is that the larger the modulus is, the smaller number of its multiples is (i. e., the better detection ability is). Those two intuitions give us a good modulus which is 127 for matrix $B$ as it is the biggest odd number in the range of $B$. In the rest of this section, we then use 127 as the modulus to simplify the calculation of detection ability. 

We quantify the detection ability in terms of probability. Specifically, the detection ability is measured by the probability our modulus based error detection method can detect error(s) when the result matrix $C$ is indeed corrupted. This metric is also known as the \emph{true positive rate}.

\subsubsection{Memory error in 8-bit matrix $ B $}
\label{sec: detection ability analysis of with mem error in B}

An error in matrix $B$ can propagate to corrupt a whole column of matrix $C$. Specifically, suppose the corruption happens at $B[i][j]$ and result in a difference of $d$. The $j$-th column of result matrix will be corrupted. Since $B[i][j]$ will be multiplied by $A[p][i]$ for all $p$ in $[0, m-1]$, the difference of the corresponding row sums in the result matrix will be $d * A[p][i]$. Recall that ABFT cannot detect soft errors if $|d * A[p][i]|$ is a multiple of $127$. Notice that $127$ is a prime number. By Euclid's lemma\footnote{If a prime number $a$ divides the product, $b*c$, $a$ divides $b$ or $c$.}, $|d * A[p][i]|$ is multiple of $127$ 
if and only if $|d|$ or $|A[p][i]|$ is a multiple of $127$. In the first fault model (random bit-flip at $B[i][j]$), $|d|$ could be $2^l$ where $l \in [0, 7]$.
$|d * A[p][i]|$ is a multiple of $127$ if and only if $|A[p][i]|$ equals 127, 254, or 0 since matrix A is in 8-bit unsigned integers. 
i. e., the $p$-th row will not detect the soft error in probability of $\frac{3}{256}$ assuming $A[p][i]$ randomly ranges in $[0, 255]$. Since all rows will be checked by ABFT, the probability that the error is not detected by all rows will be $(\frac{3}{256})^m$. Thus, the error is detected in the probability of $1-(\frac{3}{256})^m \geq 98.83\%$.

In the second fault model, $|d|$ can be random in the range of $[1, 255]$. $|d * A[p][i]|$ is multiple of $127$ if and only if $|d|$ equals 127 or $A[p][i]$ equals 127, 254, or 0. That is, the $p$-th row will not detect the soft error in probability of $$ \frac{1*256 + 255*3-3}{255*128}=\frac{1018}{32640} $$ assuming $A[p][i]$ is uniformly distributed in $[0, 255]$. Similar to the above analysis, the probability that the error is not detected by all rows will be $(\frac{1018}{32640})^m$. Thus, the error is detected with probability $1-(\frac{1018}{32640})^m \geq 96.89\%$.

\subsubsection{Memory error in 32-bit intermediate result matrix $C$ ($C_{\rm temp}$)}
\label{sec: analysis of mem err in c}
In the first fault model, a random bit-flip in $C$ implies the absolute value of difference of its corrupted row sum from its expected value to be $2^i$ for $i \in [0, 31]$. Thus, the error will be detected with probability 100\% since 127 cannot divide any $2^i$ for $i \in [0, 31]$.

In the second fault model, suppose a random element $c$ in $C$ is changed to  another arbitrary value $c'$. Then the difference in absolute value of its corrupted row sum and expected one is also $|c'-c|$. Think about $c$ is located somewhere in an interval of $[-2^{31}, 2^{31}-1]$. We can conclude the range of $|c'-c|$ is $(0, 2^{31} - 1 - c]$ or $(0, c+2^{31}]$ { where $2^{31} - 1 - c$ is taken when $c'=2^{31}-1$ and  $c+2^{31}$ is taken when $c' = -2^{31}$.}
Denote the number of multiples of $mod$ in the range of $(0, a]$ by $f(a)$. We can prove the following property, $f(a) + f(b) \leq f(a+b)$. The key is that if $mod$ divides $a$ and $b$, $f(a) + f(b) = f(a+b)$.
\begin{align*}
f(a) + f(b) &= f(a - a\%mod) + f(b - b\%mod) \\
&= f(a - a\%mod + b - b\%mod)\\
&= f(a + b - (a\%mod + b\%mod)) \\
&\leq f(a + b)
\end{align*}

Thus, number of multiples of mod in the range of $(0, 2^{31} - 1 - c]$ and $(0, c+2^{31}]$ is less than $f(2^{31} - 1 - c + c+2^{31}) = f(2^{31}-1)= \frac{2^{31}-1}{mod}$. Thus the detection probability of an error in this model will be at least
$1 - \frac{1}{mod} = 99.21\%$.


\subsubsection{Memory error in matrix $ A $ and computational error}
As we mentioned, matrix B takes much larger memory space and resides in the memory much longer than matrix A. To keep ABFT overhead low, we only encode matrix B and this means we do not provide memory error detection for matrix A. A computational soft error will corrupt the intermediate result of $A[i][k] * B[k][j]$. Thus it has the same behaviour as memory errors do in the 32-bit result matrix, $C$ where we discussed in Section~\ref{sec: analysis of mem err in c}.

\noComment{Peter}{You may want to either limit Algorithm 1 to return the $C_{\rm temp}$ only. Otherwise, you may need to include all those auxiliary information needed for requantization. Those involve the input scales, biases, AND also output scale, output bias and that we need to compute sum of the columns of $A$ and sum of the rows of $B$. Since the text above has been concentrating on detecting error in computing $C_{\rm temp}$, it may be OK to have the algorithm just for that. But up to you.}
\begin{algorithm}[h]
\caption{ABFT for low-precision GEMM}
\textbf{Input:} 8-bit integer matrix $A$, $B$; dimension sizes $m$, $n$, $k$

\textbf{Output:} 32-bit integer matrix $C_{\rm temp}$; number of corrupted rows
\begin{algorithmic}[1]
\STATE $mod$ $\gets$ 127
\FOR{$i$ from $0$ to $k-1$}
\STATE $rowSum[i] \gets \sum_{j = 0}^{n-1} {B[i][j]}$
\STATE $rowSum[i] \  \%= mod$
\ENDFOR
\STATE $packedEncodedB$ $\gets$ pack($B$, $rowSum[]$)
\STATE allocate $C_{\rm temp}[m][n+1]$
\STATE $C_{\rm temp}[][] \gets A * packedEncodedB$
\STATE $errCount \gets 0$
\FOR{$i$ from $0$ to $m-1$}
\STATE $tSum \gets \sum_{j = 0}^{n-1} {C_{\rm temp}[i][j]}$
\IF{$ tSum \% mod \neq C_{\rm temp}[i][n] \% mod$}
\STATE $errCount$++
\ENDIF
\ENDFOR
\RETURN $C_{\rm temp}$; $errCount$
\end{algorithmic}
\label{alg: abft for gemm}
\end{algorithm}

The complete look of our customized ABFT for low-precision GEMM is presented in Algorithm~\ref{alg: abft for gemm}.
\section{ABFT for low-precision EmbeddingBag}
\label{sec: abft for eb}
\subsection{ABFT for EB}

\noComment{Peter}{I think there are quite a bit of problems here. The ``$n$'' is not explained. The picture seems to suggest $n$ is the number of indices in the index set $\indset$ (note I made a change to use this calligraphic $I$ everywhere as the subscript $I$ is used previously to indicate that a matrix is of integer value). This is because the picture shows the indices to be $i_1$, $i_2$, up to $i_n$. On the other hand, this $n$ also denotes the dimension of the embedding row, that is, each row has $n$ entries. Please use $d$ for the embedding row dimension, which is what I use in the discussion in earlier sections.}

\begin{figure}[h]
    \centering
    \includegraphics[width=0.8\columnwidth]{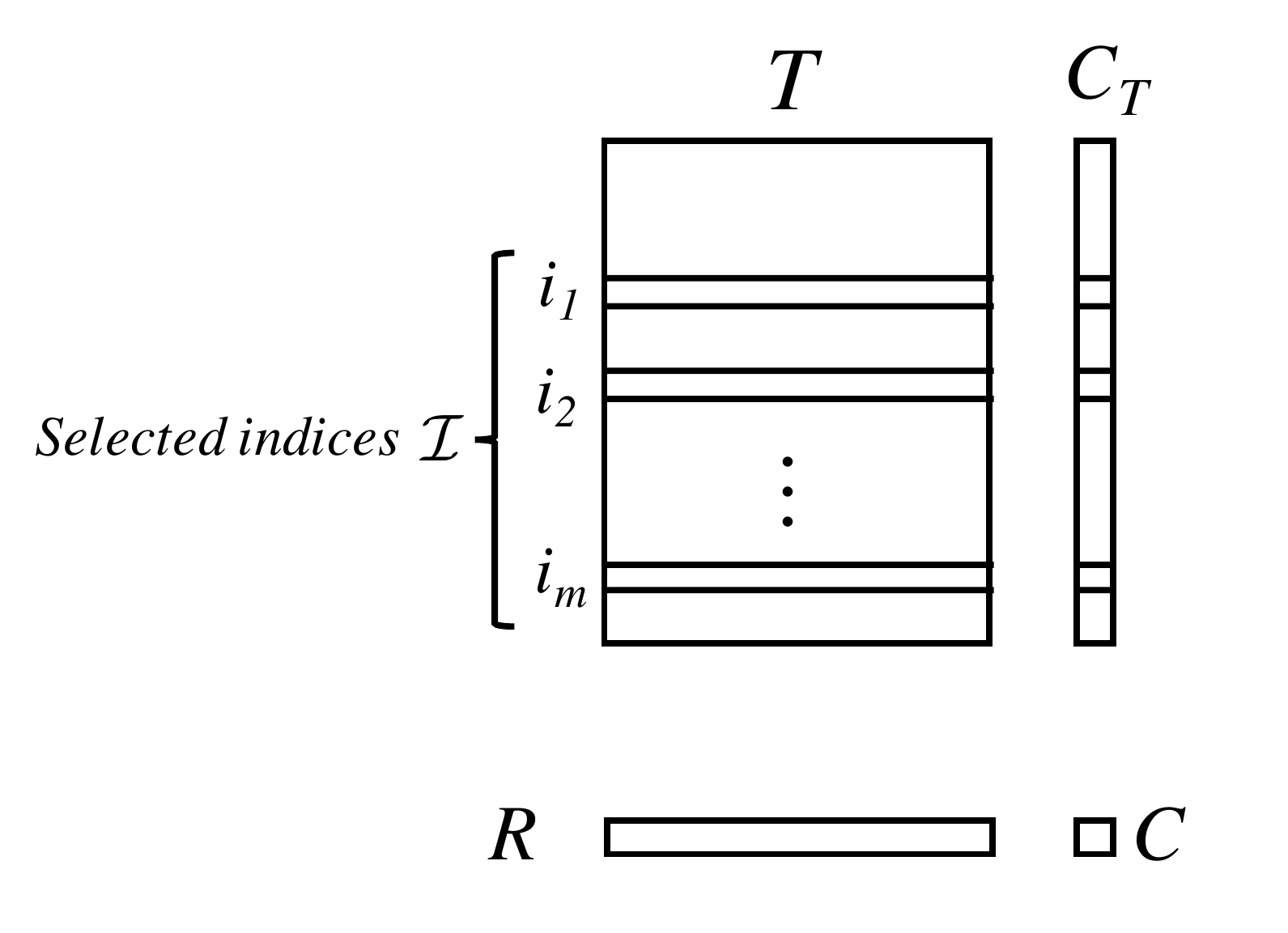}
    \caption{Illustration of ABFT for EB with batch size one}
    \label{fig: abft eb}
\end{figure}

Based on the EB operator we introduced in Section~\ref{sec: eb background}, we propose the ABFT technique for EB. To our best knowledge, this is the first ABFT technique for EB operator. { Recall that we use $d$ to denote the embedding row dimension.} The method is illustrated in Figure~\ref{fig: abft eb}.
$\vec{C_T}$ is a column vector storing all the row sums of the embedding table. If we sum the elements of $\vec{C_T}$ at the indices of $\indset$, it is easy to find the result, $C$, will be equal to the sum of all elements in $\vec{R}$. Specifically, the following equality holds. ABFT will check if this equality holds to detect soft errors.
\begin{equation}
    \sum_{j = 0}^{d-1} \vec{R}[j] = C = \sum_{i \in \indset} {\vec{C_T}[i]}
    \label{eqt: abft eb check}
\end{equation}
If the batch size is more than one, we just apply the equality check for all EBs in the batch.

\subsection{Adaption to low-precision EB}
Recall the low-precision EB variant we introduced in Section~\ref{sec: eb background}. Each embedding row vector in low-precision integers will be multiplied by a scale factor $\scale{i}$  and added by a bias value $\bias{i}$. Then equation~\ref{eqt: abft eb check} should also be updated to accommodate the scale factor and bias as shown in Equation~\ref{eqt: abft low precision eb check}.
\begin{equation}
    \sum_{j = 0}^{d-1} \vec{R}[j] = \sum_{i \in \indset} {\scale{i}*\vec{C_T}[i] + d*\bias{i}}
    \label{eqt: abft low precision eb check}
\end{equation}
The correctness of the above equation is shown as following. Recall that $\vec{e_d}$ is a $d$-length vector of all ones.
\begin{align*}
\sum_{j = 0}^{d-1} \vec{R}[j] &= \sum_{j = 0}^{d-1} {(\sum_{i \in \indset} {(\scale{i}*\vec{eb_i}[j] + \bias{i}*\vec{e_d}[j])})} \\
&=  \sum_{i \in \indset} {(\sum_{j = 0}^{d-1} {(\scale{i}*\vec{eb_i}[j] + \bias{i}*\vec{e_d}[j])})}\\
&= \sum_{i \in \indset} {(\scale{i}* \sum_{j = 0}^{d-1} {\vec{eb_i}[j]} + \sum_{j = 0}^{d-1} {\bias{i}} )} \\
&= \sum_{i \in \indset} {\scale{i}*\vec{C_T}[i] + d*\bias{i}}
\end{align*}

Notice that instead of storing the scaled and bias row sums in 32-bit float type, we still store the row sums in 32-bit integers without being scaled or biased in $\vec{C_T}$. This way we can minimize the accumulation of round off errors when we sum up the elements in $\vec{C_T}$. The details of ABFT for low-precision EB is presented in Algorithm~\ref{alg: abft for eb}.

\begin{algorithm}[h]
\caption{ABFT for low-precision EB}
\textbf{Input:} embedding table $T$; length of embedding vector, $d$; scale factor array, $\pscale[]$; bias array, $\pbias[]$; set of selected indices, $\indset$; pre-computed row sums of $T$, $C_T[]$

\textbf{Output:} EB result, $R[]$; \emph{err}
\begin{algorithmic}[1]
\STATE $R[] \gets $ EmbeddingBag($T$, $d$, $\pscale[]$, $\pbias[]$, $\indset$)
\STATE $RSum \gets \sum_{j = 0}^{d-1} {R[j]} $
\STATE $CSum \gets \sum_{i \in \indset} {(\pscale[i] * C_T[i] + d * \pbias[i])} $
\STATE $err \gets$ ``False"
\IF{$|RSum - CSum| >$ roundOffErrorBound}
\STATE $err \gets$ ``True"
\ENDIF
\RETURN $R[]$; $err$
\end{algorithmic}
\label{alg: abft for eb}
\end{algorithm}

\subsection{Overhead analysis}
Denote the number of selected indices by $m$ and the length of the embedding vector by $d$. Notice that in Algorithm~\ref{alg: abft for eb}, the row sums of embedding table is pre-computed. This can be done because once the embedding table is trained, it will stay unchanged like the weight matrix (matrix $B$) in FC layers. Thus, we do not include the operations to calculate row sums as the ABFT overhead. The number of operations in the original EB without ABFT is $3md$ and extra operations for ABFT is $3m + d$. So the overhead in fraction is $\frac{1}{d} + \frac{1}{3m}$.  In terms of memory overhead, the 32-bit row sums will take $\frac{32}{pd}$ more memory space where $p$ is the number of bits (4 or 8) of the low-precision integer in the embedding table.

\subsection{Round off error bound}
Unlike low precision GEMM where all calculations involve only integer, EmbeddingBag operators have floating point numbers where round off error can accumulate. We set a bound to differentiate soft error from round off error in $RSum$ and $CSum$ (as shown in line 5 of Algorithm~\ref{alg: abft for eb}). Setting an appropriate bound is nontrivial \cite{liang2017correcting} because too large a bound will let lots of soft error escape from the detection and too small means very high false positive rate. We choose a relative bound 1E-5 for our EmbeddingBag operators. This is a loose bound but its detection accuracy is good enough as we will show later. Why we choose a loose bound is because soft errors leading to small fluctuation of floating point results usually does not have big impact to the final machine learning inference~\cite{li2017understanding}.
\section{Evaluation}
\label{sec:evaluation}
In this section, we evaluate our proposed ABFT soft error detection for low-precision GEMM and EmbeddingBag. The solutions are evaluated in both error free case and erroneous case. A good soft error detector should have two properties: low performance overhead and low (or no) false positives in error free case; great detection ability (or high true positives) in erroneous case. 
\subsection{Performance overhead}
\subsubsection{ABFT for low-precision GEMM}
\begin{figure*}[h]
    \centering
    \includegraphics[width=0.9\textwidth]{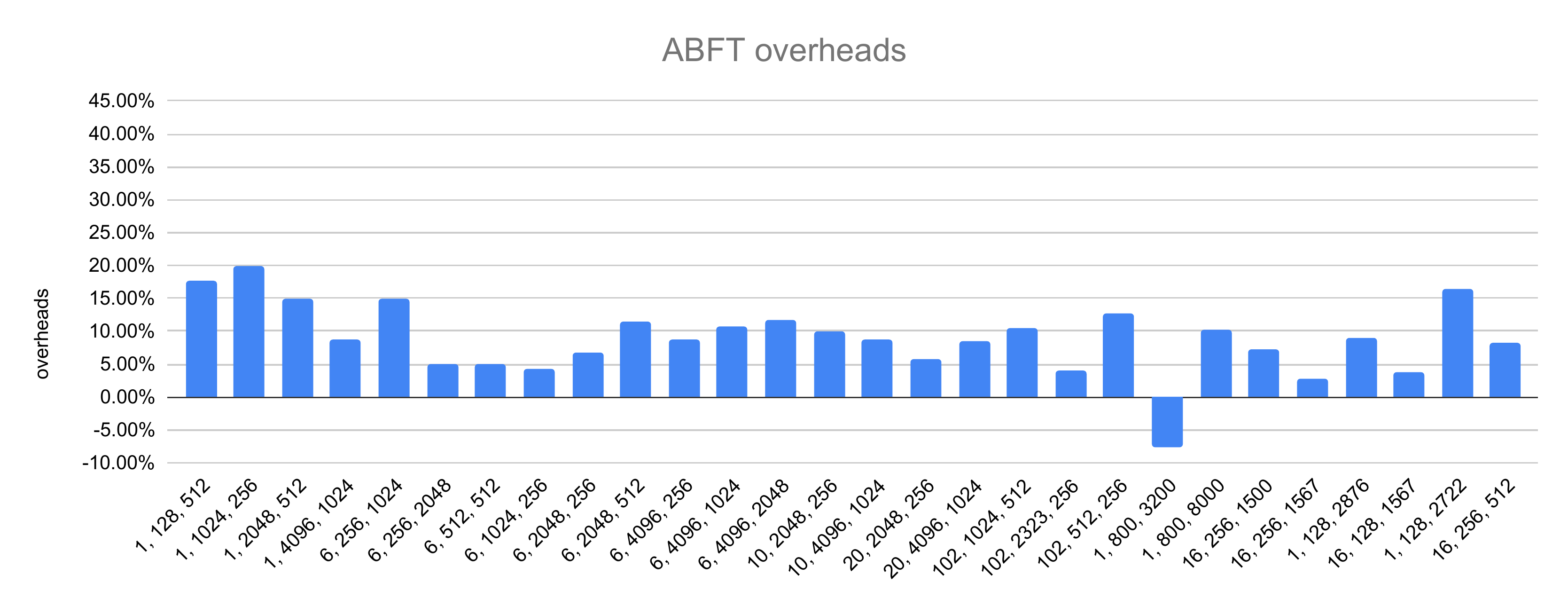}
    \caption{Performance overhead of ABFT for low-precision GEMMs with different shapes ($m, n, k$)}
    \label{fig: abft gemm ovhd}
    \vspace{-2mm}
\end{figure*}
Without any soft errors, Figure~\ref{fig: abft gemm ovhd} shows the the performance overhead of our ABFT for low-precision GEMM with different input matrix shapes. Notice that those shapes are frequently used in DLRM and they are not square. We can see from the figure that the ABFT overheads are under 20\% for all the 28 shapes. Actually, ABFT overheads are under 10\% for many of the shapes (17 out of 28 shapes); under 5\% for 7 of the shapes. Notice that for the shape $(m, n, k)$ of 1, 800, 3200, GEMM runs faster than its unprotected version. We think the reason is for that specific setting, adding one more column to matrix B makes the cache performance better.

\subsubsection{ABFT for low-precision EmbeddingBag}
We test the performance overheads of our proposed error detection method (Algorithm~\ref{alg: abft for eb}) using quantized 8-bit integer embedding table. We flush the cache since the embedding table is too large to be hold in the cache in real world scenario. We tested both regular sum and weighted sum with prefetching optimization turned on and off. The specific parameters we use is listed in Table~\ref{table: eb exp parameters}. The table columns are also known as embedding dimensions. The average pooling size is the average number of pooled embedding table rows by all EBs in a batch. For example, suppose a batch of two EBs. The first one takes 3 rows from the table and the second takes 5. The average pooling size will be 4. 

\begin{table}[h]
    \centering
    \caption{Embedding table size and experimental parameters for ABFT EmbeddingBag}
    \begin{tabular}{c|c|c|c}
    \whline
    \rowcolor[gray]{0.8}
        table rows & table columns & average pooling size & batch size \\
    \hline
          4,000,000 & 32 &100&10\\
         \hline
          4,000,000 & 64 &100&10\\
         \hline
          4,000,000 & 128 &100&10\\
         \hline
         4,000,000 & 256 &100&10\\
    \whline
    \end{tabular}
    
    \label{table: eb exp parameters}
\end{table}

\begin{figure}
    \centering
    \subfigure[without prefetching]{
    \includegraphics[width=0.9\columnwidth]{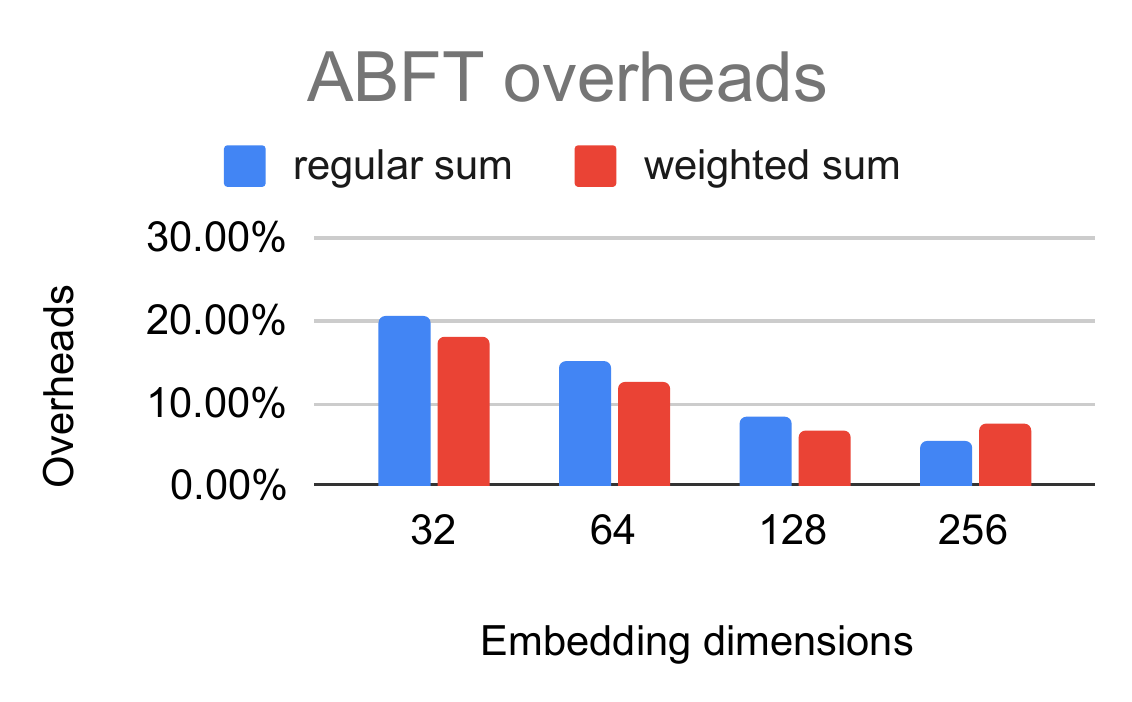}}
    \subfigure[with prefetching]{
    \includegraphics[width=0.9\columnwidth]{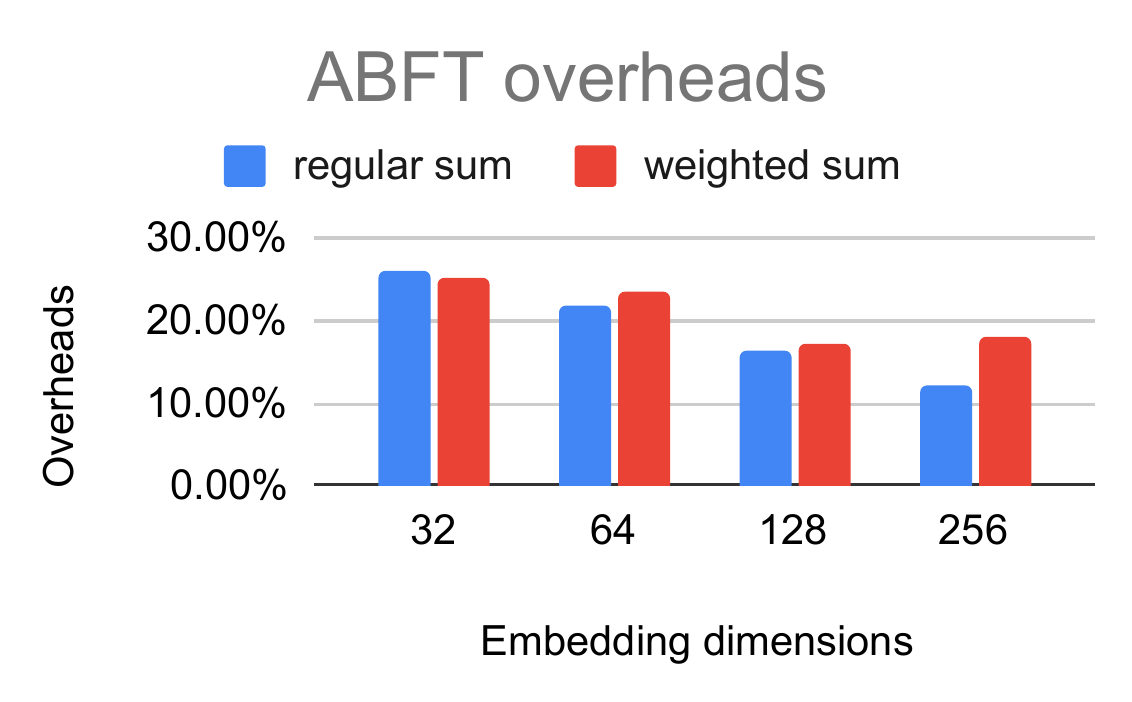}}
    
    \caption{Performance overheads of ABFT for low-precision EmbeddingBag with different settings}
    \label{fig: eb overhead exp results}
\end{figure}

\subsection{Experiments with simulated error}
We evaluate the detection accuracy of our proposed detection with simulated errors at source code level. The simulated errors are done by randomly selecting an element in the input or output and flipping a random bit in that element.

\subsubsection{ABFT for low-precision GEMM}
We first inject a random bit flip in the input matrix B after the checksum of B has been calculated and repeat the experiments for each shape 100 times totalling 2800 samples. Then we do the random bit flip injection to the 32-bit intermediate result matrix and conduct another 2800 samples. The results are shown in table~\ref{tab: sim err for gemm}.  We can see that the detection accuracy when matrix B is injected with error is $\frac{2663}{2800} = 95.11\%$. This is 3.72\% less than the theoretical estimation in Section~\ref{sec: detection ability analysis of with mem error in B} but still very high. We achieve 100\% detection accuracy when the random bit flip happens in matrix C and it is consistent with our analysis in Section~\ref{sec: analysis of mem err in c}. It is worth noting that we also conducted 2800 runs of error free experiments to validate our false positive rate is zero since there is no round off error in integer operations.

\begin{table}[t]
    \centering
    \caption{Number of detected runs and not detected runs with simulated error in GEMM}
    \begin{tabular}{>{\columncolor[gray]{0.8}} c|c|c|c}
    \whline
    \rowcolor[gray]{0.8}
         & error in B & error in C & no error\\
         \hline
    detected runs   & 2663 &2800 &0 \\
    \hline
    not detected runs & 137&0 &2800 \\
    \hline
    total &2800 &2800 &2800 \\
    \whline
    \end{tabular}
    
    \label{tab: sim err for gemm}
\end{table}

\subsubsection{ABFT for low-precision EmbeddingBag}
We tested the detection accuracy of our proposed solution with 8-bit integer embedding table. For each run, we randomly choose an element and flip a random bit in it. We repeated 400 runs with injected errors and 400 runs without injected errors. Among those 400 runs with errors, 200 of them are injected with bit flips in the upper 4 significant bits and the other 200 are injected with bit flips in the lower 4 insignificant bits. The results are shown in Table~\ref{tab: sim err for eb}. We can see the detection rate for significant 4 bits are pretty high at 99.5\%. The detection rate for insignificant 4 bits are dropped to 47\%. The false positive  rate is 9.5\%. As we can see from the results, our bound is chosen to be loose so that we can have lower false positive rates and the bad thing is that for an insignificant bit flip, detection rate is not high.

\begin{table}[t]
    \centering
    \caption{Number of detected runs and not detected runs with simulated error in EmbeddingBag}
    \begin{tabular}{>{\columncolor[gray]{0.8}} c|c|c|c}
    \whline
    \rowcolor[gray]{0.8}
         & high bits & low bits & no error\\
         \hline
    detected runs   &199 & 94& 38\\
    \hline
    not detected runs &1 &106 &362 \\
    \hline
    total &200 &200 &400 \\
    \whline
    \end{tabular}
    
    \label{tab: sim err for eb}
\end{table}

\section{Conclusion and Future Work}
\label{sec:conclusion}
In this paper, we propose efficient algorithm-based soft error detections for two important low-precision operators, GEMM and EmbeddingBag, in deep learning recommendation models. This is also the first work to benefit those deep learning operators unlike others focusing on convolutional workloads. By careful design and optimization, our proposed soft-error detection can achieve greater than 95\% in error detection ability and introduces small overheads less than 26\%.

A couple of directions we can continue to explore in the future include GPU platform migration and optimization, deployment to deep learning supercomputers to discover failure prone nodes and exploration of efficient software level error detection for other operations in DLRMs.

\bibliographystyle{IEEEtran}
\bibliography{bib/refs}

\end{document}